# Predicting the dissolution kinetics of silicate glasses using machine learning

N. M. Anoop Krishnan[1,*], Sujith Mangalathu[2], Morten M. Smedskjaer[3], Adama Tandia[4], Henry Burton[2], Mathieu Bauchy[5,*]

[1]Department of Civil Engineering, Indian Institute of Technology Delhi, Hauz Khas, New Delhi 110016, India
[2]Department of Civil and Environmental Engineering, University of California, Los Angeles, CA 90095, USA
[3]Department of Chemistry and Bioscience, Aalborg University, Aalborg, Denmark
[4]Science and Technology Division, Corning Incorporated, Corning, New York 14831, USA
[5]Physics of AmoRphous and Inorganic Solids Laboratory (PARISlab), Department of Civil and Environmental Engineering, University of California, Los Angeles, CA 90095, USA

[*]Corresponding authors: N. M. A. Krishnan (krishnan@iitd.ac.in), M. Bauchy (bauchy@ucla.edu)

## Abstract

Predicting the dissolution rates of silicate glasses in aqueous conditions is a complex task as the underlying mechanism(s) remain poorly understood and the dissolution kinetics can depend on a large number of intrinsic and extrinsic factors. Here, we assess the potential of data-driven models based on machine learning to predict the dissolution rates of various aluminosilicate glasses exposed to a wide range of solution pH values, from acidic to caustic conditions. Four classes of machine learning methods are investigated, namely, linear regression, support vector machine regression, random forest, and artificial neural network. We observe that, although linear methods all fail to describe the dissolution kinetics, the artificial neural network approach offers excellent predictions, thanks to its inherent ability to handle non-linear data. Overall, we suggest that a more extensive use of machine learning approaches could significantly accelerate the design of novel glasses with tailored properties.

## Introduction

Silicate glasses are often exposed to water—from the manufacturing stage to their service lifetime—which can result in corrosion and dissolution [1–5]. The durability of glasses in aqueous environments plays a critical role in various applications and processes, including bioactive glasses, laboratory glassware, atmospheric weathering of outdoor glasses, post-manufacturing treatment, nuclear waste immobilization, geological processes, or dissolution-precipitation-induced creep [6–16].

Depending on each application, glass dissolution may be desirable or not. As such, developing novel glasses with tailored dissolution rates requires an accurate prediction of their dissolution kinetics. However, little is known about the mechanism of silicates' dissolution and on how the dissolution kinetics depends on intrinsic (glass' composition, structure, surface geometry etc.) and extrinsic conditions (temperature, pressure, solvent chemistry, etc.) [1,13,17–20]. Elucidating composition–durability relationships in silicate glasses is further complicated by the facts that glasses can exhibit a virtually infinite number of possible compositions and that dissolution kinetics is highly non-additive with respect to composition [20–23]. These issues are further complicated by the fact that the rate-limiting mechanism of dissolution can change over time, as exemplified in the cases of silicate minerals [8] or nuclear waste glasses [10,12]. To this end, various empirical and mechanism-based models have been suggested to predict the dissolution rate of oxide glasses [1,13,17,19,24–28]. However, these models are usually applicable only for prescribed glass composition envelopes, solvent chemistry (e.g., acidic or caustic), thermodynamic conditions (e.g., temperature range). Although a mechanistic model of dissolution transferable to a broad range of glass compositions would be highly desirable, this task might not be realistic due to the structural complexity of glasses and the fact that several dissolution mechanisms can be observed, individually or in combination.

As an alternative route, data-driven models relying on machine learning are a promising tool to predict composition–property relationships in glasses based on analysis of the large quantities of experimental data that are already available [29–34]. Data-driven predictive models range from simple regression-based methods to highly non-linear methods, such as artificial neural networks—usability of which depends on the complexity of the mechanism involved [35–38]. Such methods exploit available databases of high-quality measurements to develop semi-empirical models to improve predictive capabilities [35]. These methods have been used for a wide range of applications, ranging from face recognition [39] to infrastructure lifespan prediction [40] or the design of novel composites [41]. However, very few studies have been carried out and published for the use these methods for predicting the properties of glasses [30,42,43].

Here, we investigate the ability of some machine learning approaches to predict the dissolution kinetics of a selection of sodium aluminosilicate glasses. In particular, we use linear regression (LR), random forest (RF), support vector machine regression (SVM), and artificial neural network (ANN), which represent four different classes of machine learning techniques. On account of the intrinsically non-linear character of the composition–dissolution relationship, we demonstrate that the ANN approach offers the most reliable prediction of the $SiO_2$ leaching rate over a wide range of glass compositions.

## Methodology
### Data set
To test the predictive capability of different machine learning methods, we rely on the database of dissolution rates reported by Hamilton [23]. The experiments were conducted on 9 different sodium aluminosilicate glasses, including albite glass ($Na_2O$–$Al_2O_3$–$6SiO_2$), jadeite glass ($Na_2O$—$Al_2O_3$—$4SiO_2$), nepheline glass ($Na_2O$—$Al_2O_3$—$2SiO_2$), and $Na_2O$—$xAl_2O_3$—$(3 - x)SiO_2$ glasses, where $x$ = 0.0, 0.2, 0.4, 0.6, 0.8, and 1.0. The composition range thus covers both tectosilicate and peralkaline compositions, with varying ratio of bridging to non-bridging oxygens. For each composition, the dissolution experiments were carried out in aqueous solutions with five different pH values covering both acidic and caustic conditions, specifically, pH = 1, 2, 4, 6.4, 9, and 12. For each pH, the extent of dissolution was assessed from the concentration of leached $SiO_2$ in solution after 24, 49, 96, 168, and 336 hours of solvent contact, respectively. In each case, the pH was recorded before any dissolution and at the time of the dissolution measurement. All the experiments were conducted at 25°C. For a detailed description of the measurements, the reader is invited to refer to Ref. [23].

**Inputs and outputs**
Here, our goal is to develop a predictive model of the dissolution kinetics of silicate glasses. The output of the model is chosen as being the $SiO_2$ leaching rate (in units of log[mol $SiO_2/cm^2$/s]) as this quantity captures the dissolution of the silicate skeleton of the glass. This gives a total of 299 data points. However, the methodology developed herein is general and can be applied to other outputs (e.g., the $Na_2O$ leaching rate or the glass weight loss rate). Based on the information contained in the selected database, the following variables are used as inputs: (i) the composition of the glass, (ii) the initial pH of the solution, and (iii) the pH at the time of measurement. Note that the dissolution rate was found to be fairly constant over the measured period, so that time was not included as an input. The temperature is not used since it is assumed to be constant over time.

**Machine learning methodology**
For most machine learning methods, the available data (inputs and outputs) is randomly divided into (i) a training set and (ii) a test set. The training set and test set are scattered within the whole area occupied by data due to the random sampling. The training set is first used to train the model, that is, to optimize the parameters that relate the inputs to the outputs. The test set, which is fully unknown to the model, is then used to assess the performance of the model—by comparing the outcomes of the model for inputs the model has not been explicitly trained for to reference outputs. Such division of data into training and test sets helps to avoid any potential overfitting, which is a common problem when the entire data set is used to training the model. Here, 70% and 30% of the data are attributed to the training and test sets, respectively. Note that, in the case of the ANN method, a more elaborated data classification is used, as discussed below. In the following, we provide a brief description of the predictive methodologies used herein.

*(i) Linear regression*
*(a) Simple linear regression*

The linear regression (or least squares fitting) is the simplest form of regression technique. It consists of finding the best fitting straight line through a set of points. For a given input vector $X$ = ($X_1$, $X_2$, …, $X_p$) and an output $Y$, the linear regression has the following form:

$$Y = \beta_0 + \sum_{j=1}^{p} X_j \beta_j \qquad (Eq.\ 1)$$

where $\beta_j$ are the fitting parameters of the model and $p$ the number of such parameters. The $\beta_j$ values are usually obtained by minimizing the error of predicted values with respect to the actual values, which is represented by the residual sum of squares (RSS). Thus, for a given training data set $(x_1, y_1), ..., (x_N, y_N)$ with $N$ points, the RSS can be obtained by:

$$RSS(\beta) = \sum_{i=1}^{N} (y_i - \beta_0 - \sum_{j=1}^{p} x_{ij} \beta_j)^2 \qquad (Eq.\ 2)$$

where $y_i$ is the measured value at the $i^{th}$ observation with features $x_{i,j} = (x_{i1}, x_{i2}, ..., x_{ip})$. The least square estimate of the parameters $\beta_j$ has the smallest variance among all linear unbiased estimates and, hence, is used commonly for linear regression. Note that, in unbiased estimates, all the input variables have non-zero coefficients, irrespective of whether they affect the output significantly or not.

*(b) Lasso regression*

As an unbiased estimate, one of the major drawbacks of the least square estimate is its large variance. Such variance can be reduced by introducing a small bias, wherein the unimportant input variables are neglected. To this extent, we use the lasso regression method, which typically improves the prediction accuracy of linear regression by introducing a bias and shrinking the coefficients of insignificant variables to zero. In other words, lasso regression identifies the important variables that affect the prediction significantly. To achieve this, the lasso regression method introduces a constraint on the regression coefficients using a penalty factor $\lambda$ as:

$$\hat{\beta}_{lasso} = \underset{\beta}{\mathrm{argmin}} \left\{ \sum_{i=1}^{N}(y_i - \beta_0 - \sum_{j=1}^{p} x_{ij}\beta_j)^2 + \lambda \sum_{j=1}^{p} |\beta_j| \right\} \quad \text{(Eq. 3)}$$

Note that in the lasso regression method, the penalty is imposed on $\sum_{1}^{p}|\beta_j|$.

*(c) Elastic net regression*

The variance reduction method in lasso regression can be further improved by increasing the number of constraints on the regression coefficients. To this end, Zou *et al.* [44,45] introduced a regression technique called elastic net by imposing two constraints on the regression coefficients (see Eq. 4). In the case of the elastic net regression method, the penalty is imposed on $\sum_{1}^{p}|\beta_j|$ and $\sum_{1}^{1}\beta_j^2$ and is particularly useful for analyzing high-dimensionality data. Like lasso regression, the elastic net regression method can also be used to select the significant variables. This is achieved by reducing the coefficients of irrelevant variables to zero as follows:

$$\hat{\beta}_{elastic} = \underset{\beta}{\mathrm{argmin}} \left\{ \sum_{i=1}^{N}(y_i - \beta_0 - \sum_{j=1}^{p} x_{ij}\beta_j)^2 + \frac{1-\lambda}{2}\sum_{j=1}^{p}\beta_j^2 + \lambda \sum_{j=1}^{p} |\beta_j| \right\} \quad \text{(Eq. 4)}$$

In the lasso and elastic net methods, the $\lambda$ parameter needs to be specified by the analyst. In the present study, we used a value of 0.01 based on a previous study [46].

*(ii) Support vector machine regression*

The SVM regression method reduces the error bound rather than the residual error on the training data set [37]. Hence, SVM aims to find a function $f_{SV}(x)$ that has at most a deviation $\varepsilon$ from each of the targets in the training data set. In the case of a linear function, $f_{SV}$ can be written as:

$$f_{SV}(x) = \sum_{m=1}^{M} w_m \phi_m(x) + b \quad \text{(Eq. 5)}$$

where $\phi$ is a set of $M$ mapping functions from the original data to a high-dimension feature space, $w_m$ the respective weights, and $b$ the threshold of SVM. To ensure that $w$ is as small as possible, the SVM approach reformulates the regression as an optimization problem:

$$\text{Minimize} \frac{1}{2}|w|^2$$

$$\text{subject to} \begin{cases} y_i - \sum_{m=1}^{M} w_m \phi_m(x) - b \leq \varepsilon \\ \sum_{m=1}^{M} w_m \phi_m(x) + b - y_i \leq \varepsilon \end{cases} \quad \text{(Eq. 6)}$$

The critical assumption in this formulation is that there exists a function $f_{SV}(x)$ that can approximate the training set ($x_i$, $y_i$) with a precision $\varepsilon$. To generate a strong predictive model, different mapping functions are usually used in SVM. Note that these mapping functions are augmented by kernel functions that can handle highly non-linear cases [47].

### (iii) Random forest

The RF approach relies on a collection of tree predictors. The RF method takes advantage of two powerful machine learning techniques, that is, bagging and random feature selection [38,48]. The RF approach selects a subset of features to be split at each node during the tree formation and each tree is constructed independently using a bootstrap sample of training data. The general algorithm for RF is as follows [38]:
1. Generate $n_t$ bootstrap samples from the training data set.
2. Create a decision tree from each bootstrap training sample by selecting the best split/features among the training data set. For each bootstrap iteration, predict the data that is not in the bootstrap sample—out of bag (OOB) data—using the tree grown with the bootstrap sample.
3. Predict the output of a new data set by averaging the aggregate of predictions of $n_t$ decision trees. Aggregate the OOB predictions to estimate the OOB error rate.

The output of the RF prediction can be expressed as:

$$\hat{f}_{RF}^{n_t}(x) = \frac{1}{n_t} \sum_{j=1}^{n_t} f_{RF_j}(x) \quad \text{(Eq. 7)}$$

where $\hat{f}_{RF}^{n_t}(x)$ denotes the outcome of the random forest prediction (average value) from a total of $n_t$ trees and $f_{RF}(x)$ is the individual prediction of a tree for an input vector *X*.

### (iv) Artificial neural network

The ANN algorithm is a nonlinear function of both the variables and the fitting parameters and relies on a mathematical model inspired by the behavior of biological neurons. The ANN approach consists of the input, hidden, and output layers, wherein the hidden layer contains a given number of neurons that take their inputs from the input layer and connect their outputs to the output layer. In the case of an architecture with more than one hidden layer, the outermost layer connects between the innermost and the output. Each line connecting two neurons is associated with a given weight. The output ($h_i$) of a neuron *i* in the hidden layer is calculated as:

$$h_i = s\left( \sum_{i=1}^{N} V_i x_i + T_i^{hid} \right) \quad \text{(Eq. 8)}$$

where *s*() is the activation function (or transfer function), *N* the number of input neurons, $V_i$ the weights of $i^{th}$ layer, $x_i$ the input values, and $T_i^{hid}$ the threshold term of hidden neurons. To account for the non-linearity in the composition–dissolution relationship, the activation function used herein is a sigmoid defined as [36,49,50]:

$$s(u) = \frac{1}{1+e^{-u}} \quad \text{(Eq. 9)}$$

Here, the leaching rate data are randomly split into a training set (55 % of the data), a validation set (15 % of the data), and a test set (the remaining 30 % of the data). The network is first trained with the training data by adjusting the weights between the connected neurons. The validation data set is then used to finely optimize the weight values to avoid overfitting. This is achieved by comparing the errors in the training and validation sets. For example, in the case of overfitting, although the training set may provide extremely low error values, the validation set may have a higher error value. The weight values that provide optimal errors for both training and validation sets are chosen as the final values. Finally, the efficiency of the network is estimated using the data from the test set.

## Results
### (i) Linear regression

Figure 1 shows the leaching rate values predicted by the linear regression methods selected herein. The outcomes of pure linear regression, lasso regression, and elastic net regressions are presented in Figs. 1(a), (b), and (c), respectively. In each case, the data predicted from the training and test sets are highlighted in blue and red, respectively, and the coefficient of determination $R^2$ (calculated based on either the training or test sets) is indicated. We observe that, although the measured leaching rate values range from –15 to –10 log[mol $SiO_2$/cm$^2$/s], the predicted values range only between –15 and –13 log[mol $SiO_2$/cm$^2$/s]. This suggests that the model is not able to properly describe high leaching rate data (i.e., at high pH). Further, on an individual basis, the predicted values notably differ from the measured ones. The absence of agreement between predicted and measured values is consistently observed for all the three linear regression methodologies. The inability of these regression methods to predict the leaching rate in the glasses considered herein suggests that dissolution exhibits a non-linear variation with respect to the composition and solvent pH, especially at high pH.

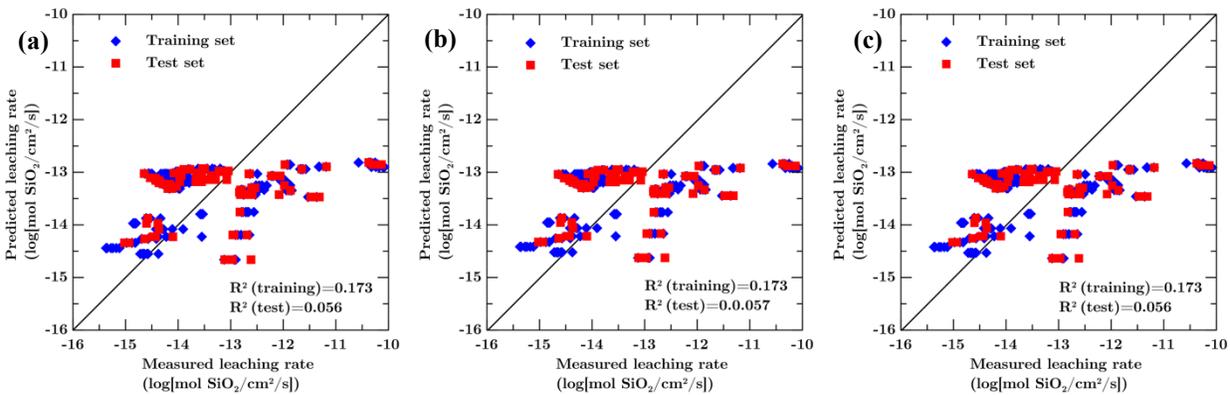

**Figure 1.** Predicted leaching rates (in log[mol $SiO_2$/cm$^2$/s]) using (a) linear regression, (b) lasso regression, and (c) elastic net, compared to the measured values.

### (ii) Support vector machine regression

The use of linear regression methods is typically limited to data sets that exhibit a linear or affine relationships between inputs and outputs. The SVM method, on the other hand, can be used to train a model based on both linear and non-linear data. Figure 2 shows the predicted leaching rate values, as compared with the measured values. Note that, here, a Gaussian kernel is used as the mapping function while training the model. Interestingly, we note that the SVM approach yields a poor prediction of the leaching rate data, similar to the case of linear regression. The discrepancy is notably high for high leaching rate data above -13 log[mol $SiO_2/cm^2$/s]. As in the case of the linear regression methods, this suggests that the model is not able to properly describe high-pH leaching rate data. More generally, we observe that the non-linear dependence of the leaching rate on glass composition and solution pH is not captured by the SVM model. This effectively limits the usability of the SVM method for predicting the dissolution rates of silicate glasses.

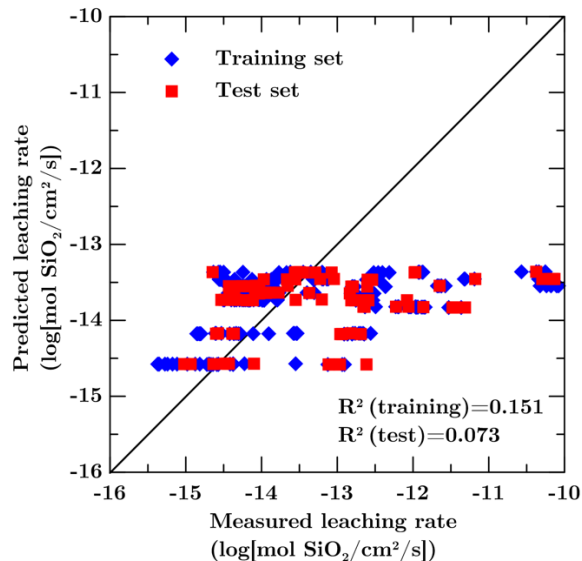

**Figure 2.** Predicted leaching rates (in log[mol $SiO_2/cm^2$/s]) using support vector regression, compared to the measured values.

### (iii) Random forest

Figure 3(a) shows the leaching rate data predicted using the RF method based on 100 trees, compared to the measured data. We observe that the RF gives a reasonably good prediction of the leaching rate data over a wide range of measured values. However, we note that the predicted values deviate from the measured ones near both of the extremes, that is, for minimum and maximum leaching rate values.

In general, the predictive accuracy of the RF approach can be improved by increasing the number of trees. However, in turn, increasing the number of trees can ultimately lead to an overfitting situation. To identify the optimum number of trees, we gradually increase the number of trees from 5 to 250. The resulting $R^2$ and mean-squared error (MSE) values of the test set are computed and plotted in Fig. 3(b). Note that, although the $R^2$ and MSE values are related to each other, they represent distinct features of the fit. Namely, $R^2$ is a measure of how close the measured data are to the predicted regression fit, i.e., the perpendicular distance between the measured data and the regression fit. However, the error corresponding to an individual prediction may not be captured, as long as they are close to the overall trend. On the other hand, the MSE value or mean squared deviation (MSD) of an estimator represents the deviation of each of the data points from the overall trend. For lower values of number of trees, the MSE of the test set decreases with increasing number of trees (Fig. 3(b)), which suggests an improved prediction. However, beyond 100 trees, we note that the MSE and $R^2$ values saturate, that is, any further increase in the number of trees results in overfitting. Thus, the optimum number of trees is here identified to be ~100. The corresponding MSE and $R^2$ values of the test set are 0.314 and 0.757, respectively.

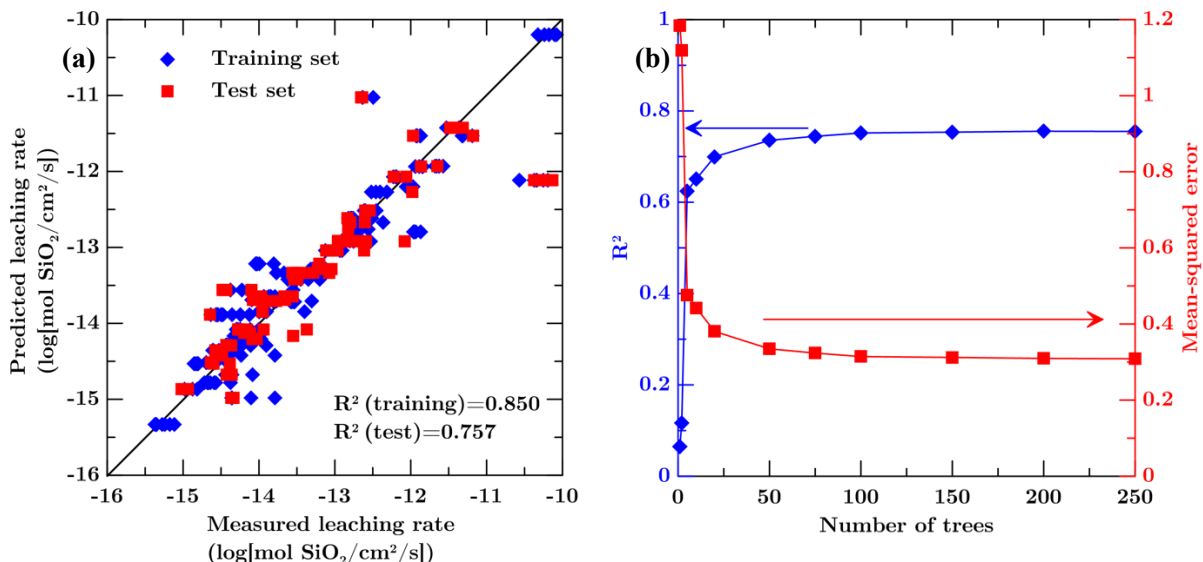

**Figure 3.** (a) Predicted leaching rates (in log[mol SiO$_2$/cm$^2$/s]) using random forest, compared to the measured values. (b) $R^2$ (left axis) and mean-squared error (right axis) values of the test set with respect to the number of trees used in the random forest algorithm.

(iv) **Artificial neural network**

The leaching rate data predicted using the ANN approach (using 10 neurons) are plotted in Fig. 4(a) and compared to the measured values. We note that the values predicted from ANN exhibit an excellent match with the measured ones over the entire range of dissolution data. In ANN, the fidelity of a model can be improved by optimizing the number of hidden layers and the number of neurons on each layer present in the network. Note that, similar to the RF method, ANN can also result in overfitting for increasing number of neurons. To optimize the network, we train the system with an increasing number of neurons, from 1 to 50 on one hidden layer. The resulting MSE and $R^2$ values of the test set are computed and shown in Fig. 4(b). We note that, for low number of neurons, the MSE initially decreases with an increase in the number of neurons. However, beyond 10 neurons, we do not observe any further decrease in the error, which suggests an optimal number of neurons of 10 for the present study. In this case, we find the MSE and $R^2$ values of the test set to be 0.027 and 0.982, respectively.

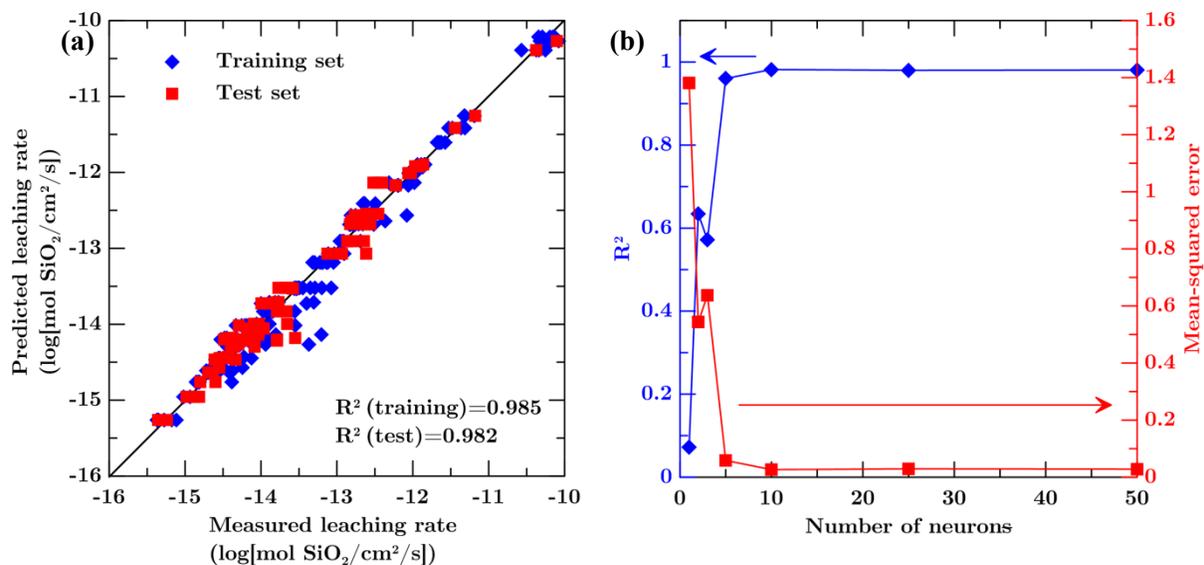

**Figure 4.** (a) Predicted leaching rates (in log[mol SiO$_2$/cm$^2$/s]) using the artificial neural network approach, compared to the measured values. (b) $R^2$ (left axis) and mean-squared error (right axis) of the test set with respect to the number of neurons in the neural network.

## Discussion

A comparison of the outcomes of the different machine learning methods presented above reveals that the ability of a given method to predict the dissolution kinetics of silicate glasses highly depends on the nature of the input-output relationship, that is, linear vs. non-linear. Traditionally, linear regressions methods can provide reasonable estimates with very low computational cost in the case of fairly linear or affine data [35]. Thus, linear regression methods can be used as a check to identify whether the input-output relationship exhibits linearity or not. In the present case, the failure of linear regression methods to predict the leaching rate data suggests that this quantity is a highly non-linear function of the glass composition and solvent pH.

Similarly, we observe that the SVM approach is also not able to yield good predictive results. Note that the success of the SVM method is highly conditioned upon the validity of the assumption that an approximation function indeed exists. Further, the predictive capability could be improved by using more appropriate kernel functions, provided that one has a preliminary understanding of the nature of the data. Here, the poor prediction of the SVM method can be attributed to the failure of the underlying assumption, as well as the usage of a simple Gaussian kernel function. Although a systematic study could be conducted to identify the appropriate kernel function for each given property, such methodology would be computationally expensive

~~and inefficient~~. In such cases, the inherent ability of the RF and ANN approaches to handle non-linear data constitutes a strong advantage.

The RF approach can be easily trained and provides reasonable predictions of the leaching rates when the number of trees is large enough. However, we note that the values that are located toward the extrema are poorly predicted by this method. This arises from the fact that RF is a model based on discrete trees and, hence, cannot be used to accurately extrapolate values beyond the range of the training set. Thus, the RF approach can provide accurate predictions only if the training set is exhaustive enough to include the entire range of the measurements. However, in many practical cases such as for the design and discovery of new glass compositions with superior properties [30], ensuring the aforementioned condition might not be feasible. Note that the inability of extrapolating far beyond the range they have been trained is in general an intrinsic limitation of data-driven mode, but different methods still exhibit different degrees of performance regarding "how far" they can extrapolate.

On the other hand, the ANN approach exhibits a significantly superior potential for extrapolation. Indeed, given the continuous non-linear functions it relies on, both extrapolation and interpolation are possible with ANN—as long as the training and test sets belong to the same class of data. Here, we observe that the ANN gives an excellent prediction of the leaching rates once trained with the training set. In addition, the developed model is able to extrapolate values beyond the range of the training set, thereby yielding accurate predictions of the dissolution rates over the entire range of values. Overall, in the context of dissolution kinetics, we observe that ANN is superior to the other machine learning methods considered herein. Extending this methodology to predict other properties of silicate glasses might require a careful and detailed analysis, as presented here, to identify the best-suited method.

Overall, these results suggest that machine learning methods can provide an efficient route toward the prediction of composition–property relationships in glasses. Glasses actually constitute an ideal class of materials for such modeling techniques since their composition can be continuously tuned thanks to the lack of stoichiometric requirements. Although it might be deceiving to rely on complex models whose linkages cannot easily be apprehended and that do not rely on any kind of underlying physics, data-driven models offer a pragmatic and reliable option while waiting for the development of mechanism-based models and improvements in computational power. However, it is worth noting that the ability of machine learning to yield accurate predictions strongly depends on the availability of large collections of reliable and consistent data, which is presently critically lacking. As such, the development and curation of databases of glass properties that are measured with consistent protocols are critical to reliable applications of such approaches.

## Conclusions

As shown in this work, the nature of the input-output relationship for prediction of glass properties—that is, whether the output (e.g., dissolution rate) depends on the input (e.g., glass composition) in a linear or non-linear fashion—can be identified using machine learning. Based on this knowledge, a suitable machine learning approach can be proposed to predict a given property, in this case, the SiO$_2$ leaching rate of silicate glasses in various conditions. Further, we observe that, although the random forest approach can reasonably predict the leaching rates, the inherent inability of the method to extrapolate data beyond the range of the training set makes it restricted to extensive databases thoroughly covering the full space of parameters. Through analysis of the outcomes of different machine learning methodologies, we demonstrate that the artificial neural network approach can be used to accurately predict the SiO$_2$ leaching rates of silicate glasses. Exploiting the abilities of such data-driven approaches can help to develop semi-empirical and physics-based models for composition–property predictions. Ultimately, this can accelerate the design of novel commercial glasses with tailored functionalities.

## Acknowledgments

The authors acknowledge some financial support by the National Science Foundation under Grant No. 1562066 and the Department of Science and Technology, India under the INSPIRE faculty scheme (DST/INSPIRE/04/2016/002774).